\documentclass{moriond}

\usepackage{amsmath,amsfonts} 
\usepackage{array}
\newcolumntype{P}[1]{>{\centering\arraybackslash}p{#1}}

\def\be{\begin{equation}}
\def\ee{\end{equation}}
\def\bea{\begin{eqnarray}}
\def\eea{\end{eqnarray}}

\newcommand{\Photo}{}

\begin{document}
\vspace*{4cm}
\title{Constraints on light scalars in the Aligned-two-Higgs-doublet model}

\author{ V. Miralles,$^{1,\,}$\footnote{Speaker and corresponding author: victor.miralles@manchester.ac.uk} A. M. Coutinho,$^2$ A. Karan,$^2$ A. Pich$^2$ }

\address{$^1$Department of Physics and Astronomy, University of Manchester, Oxford Road, Manchester M13 9PL, United Kingdom\newline
$^2$Instituto de F\'isica Corpuscular, Universitat de Val\`encia – CSIC, Parque Cient\'ifico, Catedr\'atico Jos\'e Beltr\'an 2, E-46980 Paterna, Spain}

\maketitle\abstracts{
In this work, we explore the viability of light-scalar scenarios within the $\mathcal{CP}$-conserving, flavour-aligned two-Higgs-doublet model (A2HDM), where at least one additional neutral or charged scalar is assumed to be lighter than the 125~GeV Higgs boson. The A2HDM naturally avoids flavour-changing neutral currents via Yukawa alignment, providing a more flexible framework than conventional $Z_2$-symmetric models. Employing the \texttt{HEPfit} package within a Bayesian statistical approach, we perform global fits across all distinct light-scalar configurations, incorporating theoretical requirements and experimental constraints from electroweak precision observables, flavour data, Higgs measurements, and direct searches at the LEP and the LHC.
 }

\section{The aligned-two-Higgs-doublet model}

A straightforward extension of the Standard Model (SM) scalar sector involves introducing a second complex $SU(2)_L$ doublet with hypercharge $Y=1/2$, giving rise to the two-Higgs-doublet model (2HDM). The general scalar potential consistent with $SU(2)_L \times U(1)_Y$ gauge invariance allows both doublets to acquire vacuum expectation values (VEVs), but a global $SU(2)$ transformation enables a basis—the so-called Higgs basis—where only one of them develops a VEV. 

In this basis, the scalar doublets are expressed as:
\begin{equation}
\Phi_1 = \frac{1}{\sqrt{2}} \begin{pmatrix}
\sqrt{2} G^+ \\
v + S_1 + i G^0
\end{pmatrix}, \quad
\Phi_2 = \frac{1}{\sqrt{2}} \begin{pmatrix}
\sqrt{2} H^+ \\
S_2 + i S_3
\end{pmatrix},
\end{equation}
where $v = 246$ GeV is the SM VEV, and $G^\pm,\, G^0$ are the would-be Goldstone bosons. The physical spectrum includes one charged scalar $H^\pm$ and three neutral scalars $S_i$ ($i = 1,2,3$).

The most general renormalisable scalar potential in the Higgs basis reads:
\begin{align}
\label{eq:pot}
V&=\mu_1\,\Phi_1^\dagger \Phi_1+\mu_2\,\Phi_2^\dagger \Phi_2+ \Big[\mu_3\,\Phi_1^\dagger \Phi_2+\mathrm{h.c.}\Big]+\frac{\lambda_1}{2}\,(\Phi_1^\dagger \Phi_1)^2+\frac{\lambda_2}{2}\,(\Phi_2^\dagger \Phi_2)^2+\lambda_3\,(\Phi_1^\dagger \Phi_1)(\Phi_2^\dagger \Phi_2)\nonumber\\
&+\lambda_4\,(\Phi_1^\dagger \Phi_2)(\Phi_2^\dagger \Phi_1)+\Big[\Big(\frac{\lambda_5}{2}\,\Phi_1^\dagger \Phi_2+\lambda_6 \,\Phi_1^\dagger \Phi_1 +\lambda_7 \,\Phi_2^\dagger \Phi_2\Big)(\Phi_1^\dagger \Phi_2)+ \mathrm{h.c.}\Big]\, ,
\end{align}
with $\mu_3$, $\lambda_{5,6,7}$ generally complex, although they are taken real in the $\mathcal{CP}$-conserving limit considered here.

After electroweak symmetry breaking, the neutral scalars mix. In the $\mathcal{CP}$-conserving scenario, $S_3$ remains a $\mathcal{CP}$-odd physical mass eigenstate ($A = S_3$), while the $\mathcal{CP}$-even scalars $(S_1, S_2)$ mix via a rotation:
\begin{equation}
\begin{pmatrix}
h \\
H
\end{pmatrix}
= \begin{pmatrix}
\cos\tilde{\alpha} & \sin\tilde{\alpha} \\
- \sin\tilde{\alpha} & \cos\tilde{\alpha}
\end{pmatrix}
\begin{pmatrix}
S_1 \\
S_2
\end{pmatrix},
\end{equation}
where $h$ denotes the observed 125 GeV Higgs boson, and $H$ is the beyond-SM (BSM) $\mathcal{CP}$-even state.

The potential parameters can be reparametrised in terms of physical masses and mixing angles. For instance, the mixing angle is related to the parameters of the scalar potential as:
\begin{equation}
\tan\tilde{\alpha} =\frac{M_h^2-v^2\lambda_1}{v^2\lambda_6} =\frac{v^2 \lambda_6}{v^2 \lambda_1 - M_H^2} \, .
\end{equation}

Therefore we can select the parameters of our model to be $\{$$v$, $M_{h}$, $M_{H^\pm}$, $M_{H}$, $M_{A}$, $\tilde{\alpha}$, $\lambda_2$, $\lambda_3$, $\lambda_7$$\}$, where $v$ and $M_{h}$ have already been precisely measured experimentally. For more details we refer the reader to Ref.~\cite{Coutinho:2024zyp}.

The Yukawa Lagrangian in the Higgs basis, written in the mass-eigenstate basis of the fermions, is given by:
\begin{align}
\hspace*{-5 mm}
-\mathcal{L}_Y &= \frac{\sqrt{2}}{v} \left[ \bar{Q}_L \left( M_u \tilde{\Phi}_1 + Y_u \tilde{\Phi}_2 \right) u_R + \bar{Q}_L \left( M_d \Phi_1 + Y_d \Phi_2 \right) d_R + \bar{L}_L \left( M_\ell \Phi_1 + Y_\ell \Phi_2 \right) \ell_R \right] + \text{h.c.},\hspace*{-3 mm}
\end{align}
where $M_f$ are the diagonal fermion mass matrices and $Y_f$ are general $3\times3$ matrices. In the generic 2HDM, this structure generates tree-level flavour-changing neutral currents (FCNCs), tightly constrained by experiment.
To avoid these FCNCs without imposing discrete symmetries, the A2HDM introduces an alignment condition in flavour space:
\begin{equation}
Y_f = \varsigma_f M_f, \quad \text{with} \quad \varsigma_f \in \mathbb{R},
\quad f = u, d, \ell,
\end{equation}
which eliminates off-diagonal neutral currents at tree level. The alignment parameters $\varsigma_f$ are, in general, complex, but are taken real here to ensure $\mathcal{CP}$ conservation.

In terms of the physical scalars, the Yukawa interactions become:
\begin{align}
-\mathcal{L}_Y \supset \sum_{i,f} \frac{y^{\varphi_i^0}_f}{v} \varphi_i^0 \bar{f} M_f P_R f
+ \frac{\sqrt{2}}{v} H^+ \left[ \bar{u} \left( \varsigma_d V M_d P_R - \varsigma_u M_u^\dagger V P_L \right) d + \varsigma_\ell \bar{\nu} M_\ell P_R \ell \right] + \text{h.c.},
\end{align}
where $\varphi_i^0 \in \{h, H, A\}$ and the couplings are given by:
\begin{equation}
y_f^h = \cos\tilde{\alpha} + \varsigma_f \sin\tilde{\alpha}\, ,\quad
y_f^H = -\sin\tilde{\alpha} + \varsigma_f \cos\tilde{\alpha}\, ,\quad
y_u^A = -i \varsigma_u\, ,\quad
y_{d,\ell}^A = i \varsigma_{d,\ell}\, .
\end{equation}



\section{Constraints}
\label{sec:Constraints}

In this section, we describe the different theoretical and experimental constraints imposed in our analysis. 

\subsection{Theoretical constraints}

The requirement that the scalar potential remains bounded from below for any direction in field space is crucial to guarantee a stable electroweak vacuum. In the bilinear formalism~\cite{Ivanov:2006yq}, the potential can be rewritten as
\begin{equation}
V = - M_\mu r^\mu + \frac{1}{2} \Lambda^\mu_{\ \nu} r_\mu r^\nu,
\end{equation}
where \(M_\mu\) and \(r^\mu\) are four-vectors constructed from the scalar fields,
\begin{align}
M_\mu &= \left( -\frac{1}{2}(\mu_1 + \mu_2),\ -\text{Re}(\mu_3),\ \text{Im}(\mu_3),\ -\frac{1}{2}(\mu_1 - \mu_2) \right), \\
r^\mu &= \left( |\Phi_1|^2 + |\Phi_2|^2,\ 2\,\text{Re}(\Phi_1^\dagger\Phi_2),\ 2\,\text{Im}(\Phi_1^\dagger\Phi_2),\ |\Phi_1|^2 - |\Phi_2|^2 \right),
\end{align}
and \(\Lambda^\mu_{\ \nu}\) is a real, symmetric matrix built from the quartic couplings.

The necessary and sufficient conditions for the potential to be bounded from below are: (1) the matrix \(\Lambda^\mu_{\ \nu}\) must be diagonalisable by an \(SO(1,3)\) transformation; (2) all its eigenvalues must be real; and (3) the timelike eigenvalue must be positive and larger than the spacelike ones, \(\Lambda_0 > \Lambda_{1,2,3}\).
This last condition ensures that the quartic part of the potential is positive along any direction in field space, thus guaranteeing vacuum stability.

To ensure that the vacuum corresponds to a global minimum and not merely a local one, we impose an additional constraint on the scalar mass parameters. Defining the determinant of \((\xi \mathbb{I}_4 - \Lambda^\mu_{\ \nu})\) as \(D(\xi) = -\prod_{k=0}^3 (\xi - \Lambda_k)\), and choosing \(\xi = M_{H^\pm}^2/v^2\), the global minimum condition requires either 
$
D(\xi) > 0, \; \text{or} \; (D(\xi) < 0 \; \text{and} \; \xi > \Lambda_0).
$

At high energies, the validity of the perturbative expansion imposes strong restrictions on scattering amplitudes \cite{Ginzburg:2005dt}. In particular, the scalar quartic couplings must be such that all \(2\to2\) scattering processes involving scalar states remain unitary. This condition is typically studied by expanding the scattering amplitude in partial waves. The leading $S$-wave component dominates at high energies, and its matrix form is given by:
$
(\mathbf{a}_0)_{i \to f} = \frac{1}{16\pi s} \int_{-s}^{0} dt \; \mathcal{M}_{i \to f}(s,t),
$
where \(\mathcal{M}_{i \to f}(s,t)\) is the tree-level amplitude for the transition from the initial state \(i\) to the final state \(f\). 
We diagonalise this matrix in subspaces of definite hypercharge and weak isospin, and require that all eigenvalues \(e_i\) of the resulting block matrices respect the perturbative unitarity condition:
$
|e_i| \leq 8\pi.
$
These bounds apply to all neutral, singly-charged, and doubly-charged scalar scattering channels, and result in upper limits on various combinations of the quartic couplings \(\lambda_i\). 
%
%
Finally, to preserve perturbativity in the Yukawa sector, we impose that the alignment parameters satisfy
$
|\varsigma_f| < \frac{v}{\sqrt{2} m_f},
$
which ensures that the effective scalar–fermion couplings remain perturbative.

\subsection{Electroweak precision observables}

The presence of additional scalars impacts the electroweak precision observables through their loop contributions to the vacuum polarisation of gauge bosons. These effects are encapsulated in the oblique parameters $S$ and $T$. On top of these, we also include $R_b$ (the ratio of partial widths for $Z\to b\bar{b}$) in our analysis. Note also that we have repeated the electroweak fit removing the observables that could be affected by the BSM scalars to obtain the values of $S$ and $T$ used in our fit, as described in Ref.~\cite{Karan:2023kyj}. The included electroweak observables are especially powerful in constraining the mass differences between the additional scalars.

\subsection{Flavour observables}

Low-energy flavour processes provide some of the most stringent constraints on new physics models. In the A2HDM, the new scalar particles can contribute both at tree and loop level to various flavour-changing processes \cite{Jung:2010ik}. Since the CKM matrix is determined experimentally with high precision, we treat the four Wolfenstein parameters as nuisance parameters in our fit. To avoid contamination from potential new physics contributions, we follow the procedure outlined in Ref.~\cite{Karan:2023kyj}, fitting the CKM parameters using only observables that are robust against new scalar effects.

The flavour observables here considered receiving contributions from the A2HDM at loop level are the mass difference in the $B_s$ system ($\Delta M_{B_s}$), the branching ratio for the radiative decay $B\to X_s\gamma$, and the rare leptonic decay $B_s\to\mu^+\mu^-$. We also include tree-level processes in the A2HDM such as the purely leptonic decays $B\to\tau\nu$, $D_{(s)}\to\mu\nu$, and $D_{(s)}\to\tau\nu$, where the charged scalars provide the tree-level contributions. Ratios of kaon and tau decays, like $\Gamma(K\to\mu\nu)/\Gamma(\pi\to\mu\nu)$ and $\Gamma(\tau\to K\nu)/\Gamma(\tau\to\pi\nu)$, provide additional information. These processes are particularly sensitive to the values of the alignment parameters $\varsigma_u$, $\varsigma_d$, and $\varsigma_\ell$, and impose significant restrictions on the charged Higgs mass and couplings.

The anomalous magnetic moment of the muon has not been included in the global fit but an extensive discussion on its effects is given in Ref.~\cite{Coutinho:2024zyp}. The contributions of the A2HDM with $\mathcal{CP}$-violation to the electric dipole moment of the electron has recently been instudied in Ref.~\cite{Davila:2025goc}.

\subsection{Higgs signal strengths}

The observed properties of the 125~GeV Higgs boson offer another sensitive probe of extended scalar sectors. In the A2HDM, the mixing between the two $\mathcal{CP}$-even neutral scalars modifies the couplings of the 125~GeV Higgs boson to SM fermions and gauge bosons. Furthermore, the presence of the charged Higgs alters the one-loop decay $h\to\gamma\gamma$. We include in our fit the latest signal strength measurements from ATLAS and CMS at both 8~TeV and 13~TeV, covering the dominant production modes (gluon fusion, vector boson fusion, associated production with vector bosons and with top quarks) and decay channels ($b\bar{b}$, $\gamma\gamma$, $\tau^+\tau^-$, $WW$, $ZZ$, $\mu^+\mu^-$, and $c\bar{c}$). These measurements put strong bounds on the mixing angle $\tilde{\alpha}$ and the alignment parameters, since they determine the overall rescaling of the Higgs couplings.

\subsection{Direct searches for new scalars}

Direct collider searches for additional scalar particles provide crucial information on the viability of the model. We incorporate constraints from the LEP and the LHC by comparing the theoretical predictions for production cross sections times branching ratios with the corresponding experimental upper limits. We use the packages \texttt{MadGraph5\_aMC@NLO}~\cite{Alwall:2014hca}, \texttt{HIGLU}~\cite{Spira:1995mt}, and \texttt{HDECAY}~\cite{Djouadi:2018xqq} to obtain the theoretical predictions. 

From the LEP, we consider Higgsstrahlung processes and scalar pair production. From the LHC, we take into account the resonant production of heavy neutral scalars decaying into gauge boson pairs, Higgs boson pairs, or fermion pairs, as well as charged Higgs searches in both the low- and high-mass regimes. Additionally, constraints from invisible decay widths of the Higgs, $W$, and $Z$ bosons, and the total width of the top quark, are included. These direct searches strongly restrict the masses and couplings of light scalars.

\section{Global fits}

The statistical analysis uses the open-source framework \texttt{HEPfit}~\cite{DeBlas:2019ehy}, which implements a Markov Chain Monte Carlo sampling technique via the \texttt{Bayesian Analysis Toolkit}~\cite{Caldwell:2008fw}. 

The $\mathcal{CP}$-conserving A2HDM introduces ten free BSM parameters: $\{$$M_H$, $ M_A$, $M_{H^\pm}$, $\tilde{\alpha}$, $\lambda_2$, $\lambda_3$, $\lambda_7$, $\varsigma_u$, $\varsigma_d$, $\varsigma_\ell$$\}$. For the global fit, we use uniform (flat) priors for all these parameters, with ranges chosen to accommodate the full set of theoretically and phenomenologically viable scenarios. These bounds are informed by perturbativity, vacuum stability, and compatibility with existing experimental constraints.

The scalar mass parameters are divided into two categories: scalars lighter than the 125~GeV Higgs boson (denoted \(\phi_{\text{light}}\)), which are allowed in the range \(10~\text{GeV} \leq M_{\phi_{\text{light}}} < M_h\); and heavier scalars \(\phi_{\text{heavy}}\), with masses between \(M_h\) and 700~GeV. The ranges for the couplings and alignment parameters are displayed in Tab.~\ref{tab:prior}.

\begin{table}[t!]
    \centering
\begin{tabular}{P{1.1cm}|P{1.1cm}|P{1.1cm}|P{1.1cm}|P{1.1cm}|P{1.1cm}|P{1.1cm}|P{1.1cm}|P{1.1cm}|P{1.1cm}|P{1.1cm}|P{1.1cm}|}
\hline
\multicolumn{12}{|c|}{\bf Priors} \\
\hline
\hline
\multicolumn{6}{|P{6.6 cm}}{$M_{\phi_{\text{light}}^{}} \in [10 \text{ GeV},\; M_h]$ } & \multicolumn{6}{|P{6.6 cm}|}{$M_{\phi_{\text{heavy}}^{}} \in [M_h,\; 700 \text{ GeV}]$ }  \\
\cline{2-12}
\multicolumn{4}{|P{4.4 cm}}{$\lambda_2\in [-1,\;10]$} & \multicolumn{4}{|P{4.4 cm}|}{$\lambda_3\in [-1,\;10]$}  & \multicolumn{4}{P{4.4 cm}|}{$\lambda_7\in [-3.5,\;3.5]$}  \\
\cline{2-12}
\multicolumn{3}{|P{3.3 cm}}{$\tilde \alpha\in [-0.2,\;0.2]$} & \multicolumn{3}{|P{3.3 cm}|}{$\varsigma_u\in [-0.5,\; 0.5]$} & \multicolumn{3}{P{3.3 cm}|}{$\varsigma_d \in [-10,\; 10]$} & \multicolumn{3}{P{3.3 cm}|}{$\varsigma_\ell \in [-100,\; 100]$} \\  
\hline
\end{tabular}
\caption{Flat priors used for the parameters of the A2HDM. Here, $\phi_{\text{light}}$ and $\phi_{\text{heavy}}$ refer to scalar states lighter or heavier than the SM-like Higgs, respectively, with $\{\phi\} \in \{H, A, H^\pm\}$.}
\label{tab:prior}
\end{table}

The structure of the parameter space allows for multiple configurations where one, two, or all three new scalars lie below the Higgs mass. We therefore distinguish and separately explore seven phenomenologically distinct scenarios, corresponding to all possible combinations of light scalar states. In each case, we impose the same prior ranges shown in Tab.~\ref{tab:prior}, with the additional condition that at least one BSM scalar lies below the 125~GeV threshold. 

\begin{figure}
    \centering
    \includegraphics[width=0.7\linewidth]{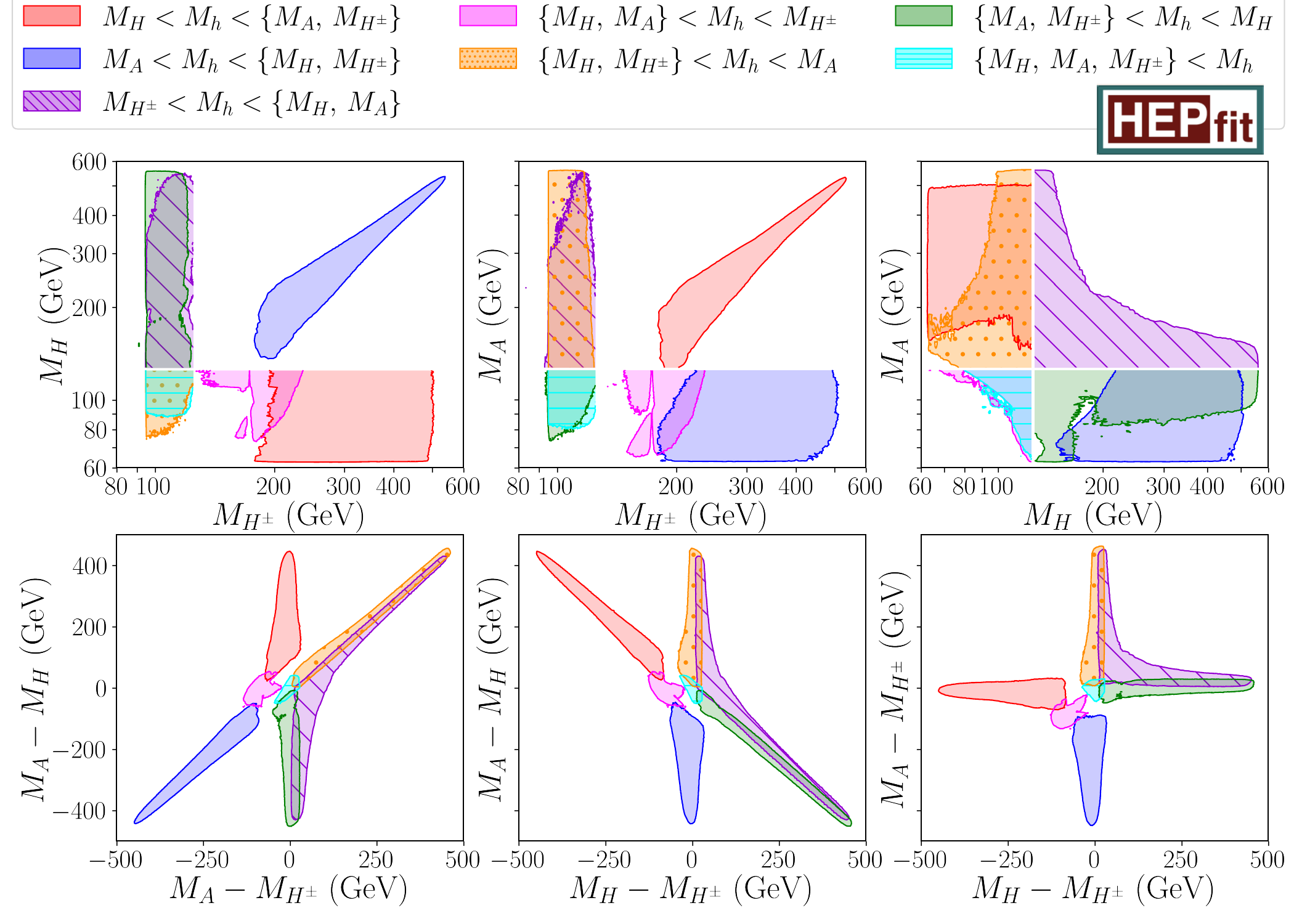}
    \caption{Correlations among the scalar masses and their mass splitting, shown as 95\% allowed probability regions.}
    \label{fig:masses}
\end{figure}

Fig.~\ref{fig:masses} displays the correlations between the scalar masses, as well as their mass splittings, for the various light-scalar scenarios considered. A clear interplay emerges, particularly between the two neutral states, driven by the requirement of compatibility with electroweak precision constraints, which impose that at least one of the neutral scalars must be close in mass to the charged scalar.

The theoretical constraints place an upper bound on the allowed mass differences among the scalar states. In particular, when one of the scalars is lighter than the 125 GeV Higgs boson, configurations with scalar masses exceeding 600~GeV become increasingly disfavoured. Complementary to this, collider searches impose strict lower limits: experimental bounds require the mass of the charged scalar to be greater than 90~GeV, while neutral scalar masses are constrained to lie above roughly 60~GeV.

\begin{figure}
    \centering
    \includegraphics[width=0.49\linewidth]{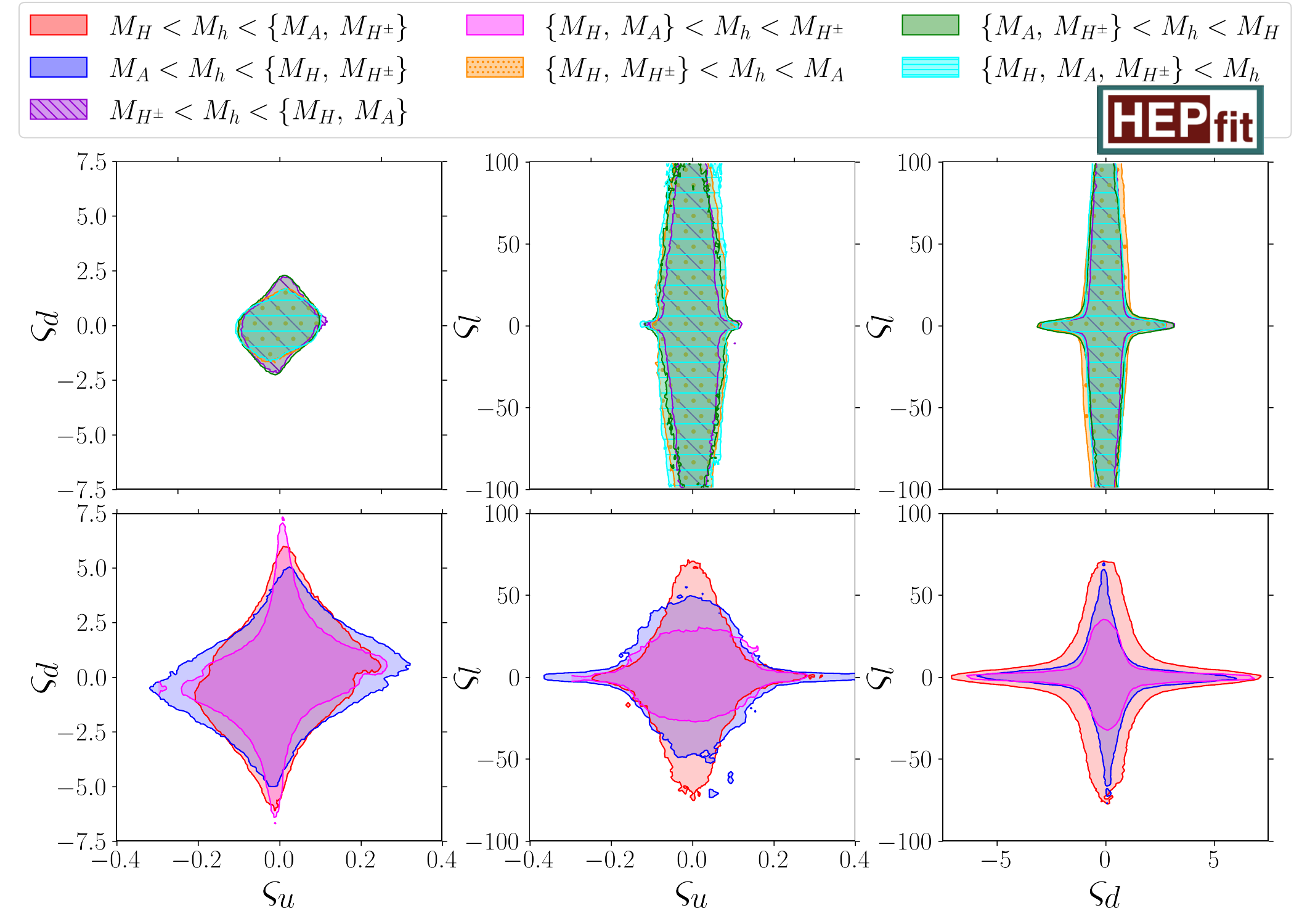}
    \includegraphics[width=0.49\linewidth]{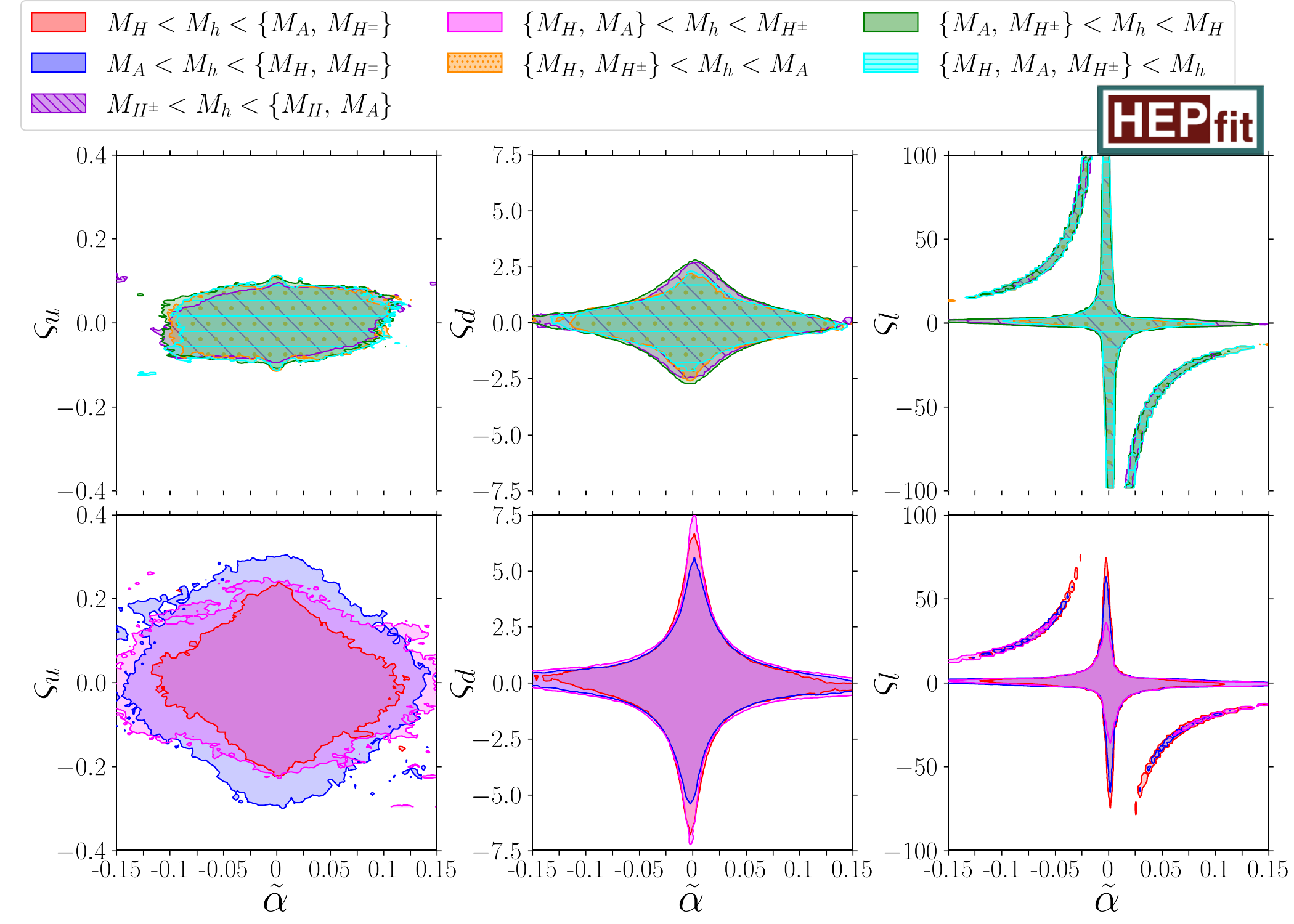}
    \caption{Correlations among the mixing angle (in rad) and the Yukawa alignment parameters, shown as 95\% allowed probability regions.}
    \label{fig:Couplings_vs_alpha}
\end{figure}

Fig.~\ref{fig:Couplings_vs_alpha} shows the correlations between the alignment parameters and the mixing angle. In scenarios where the charged scalar is light, both $\varsigma_u$ and $\varsigma_d$ are tightly constrained, whereas $\varsigma_\ell$ remains relatively unconstrained. This reflects the fact that quark-flavour observables dominate the fit and suppress any indirect sensitivity to the lepton sector. Nevertheless, sizable values of $\varsigma_\ell$ (up to $\mathcal{O}(100)$) are only possible when the quark alignment parameters are close to zero. Conversely, when $\varsigma_{u,d}$ are allowed to be moderately large, the viable range of $\varsigma_\ell$ becomes significantly narrower.

The scalar mixing angle $\tilde{\alpha}$ also exhibits a strong interplay with the Yukawa alignment parameters $\varsigma_f$. While measurements of the 125~GeV Higgs boson interacting with gauge bosons constrain $\cos\tilde{\alpha}$, the Yukawa couplings, containing a dependence proportional to $\varsigma_f \sin\tilde{\alpha}$, impose additional bounds on the mixing angle---particularly when $\varsigma_f$ is sizeable. In the limit of small alignment parameters, the 125~GeV Higgs boson Yukawa couplings approach their SM values. Regarding the correlation, since the up-type alignment parameter is already constrained by flavour data, a broad correlation among $\varsigma_u$ and $\tilde{\alpha}$ is observed, whereas tighter dependencies appear for $\varsigma_d$ and $\varsigma_\ell$. In the leptonic sector, specific regions allow a destructive interference between the two terms in the coupling, leading to viable flipped-sign solutions. These are absent for quark Yukawas due to the presence of complementary constraints. Moreover, scenarios with heavier charged scalars tend to favour wider ranges for $\varsigma_{u,d}$ and more restrictive values for $\varsigma_\ell$.

\section*{Acknowledgments}

Work supported 
by MCIU/AEI/10.13039/501100011033 (grants PID2020-114473GB-I00, PID2023-146220NB-I00 and CEX2023-001292-S), by the European Research Council under the European Union’s Horizon 2020 research and innovation programme (grant agreement 949451), and by a Royal Society University Research Fellowship (grant URF/R1/201553)

\section*{References}
\bibliography{moriond}


\end{document}